\documentclass[fleqn,twoside]{article}
\usepackage{gc,amssymb,cite}
\usepackage{latexsym}

\eqsection

\newcommand{\N}{ {\mathbb N} }
\newcommand{\R}{ {\mathbb R} }

\newcommand{\tri}{\Delta}

\begin{document}
\twocolumn[

\Title {Fluxbrane and S-brane solutions with polynomials\yy
        related to rank-2 Lie algebras}

\Aunames{I.S. Goncharenko\auth{1,a},
         V.D. Ivashchuk\auth{2,b,c} and V.N. Melnikov\auth{3,b,c}}

\Addresses{
 \addr a {School of Natural Sciences, UC Merced, 5200
    North Lake Road, Merced, CA 95344, USA}
 \addr b {Centre for Gravitation and Fundamental Metrology,
          VNIIMS, 46 Ozyornaya St., Moscow 119361, Russia}
 \addr c {Institute of Gravitation and Cosmology,
          Peoples' Friendship University of Russia,
          6 Miklukho-Maklaya St., Moscow 117198, Russia} }


\Abstract
{Composite fluxbrane and $S$-brane solutions for a wide class of
 intersection rules are considered. These solutions are defined on a
 product manifold $R_{*} \times M_1 \times \ldots \times M_n$ which contains
 $n$ Ricci-flat spaces $M_1, \ldots,  M_n$ with 1-dimensional factor spaces
 $R_{*}$ and $M_1$. They are determined up to a set of functions obeying
 non-linear differential equations equivalent to Toda-type equations with
 certain boundary conditions imposed. Exact solutions corresponding to
 configurations with two branes and intersections related to simple Lie
 algebras $C_2$ and $G_2$ are obtained. In these cases, the functions
 $H_s(z)$, $s =1,2$, are polynomials of degrees $(3,\ 4)$ and $(6,\ 10)$,
 respectively, in agreement with a conjecture put forward previously in
 Ref.\,\cite{Iflux}. The $S$-brane solutions under consideration, for
 special choices of the parameters, may describe an accelerating expansion
 of our 3-dimensional space and a small enough variation of the
 effective gravitational constant.  }


]
\email 1 {igorg@mail.ru}
\email 2 {rusgs@phys.msu.ru}
\email 3 {melnikov@phys.msu.ru}

\section{Introduction}

  In this paper, we deal with the so-called multidimensional fluxbrane
  solutions (see [1, 3--18] and references therein) that are in fact
  generalizations of the well-known Melvin solution \cite{Melv}. (Melvin's
  original solution describes the gravitational field of a magnetic
  flux tube.)

  In \cite{Iflux}, a subclass of generalized fluxbrane solutions was
  obtained. These fluxbrane solutions are governed by functions $H_s(z) > 0$
  defined on the interval $(0, +\infty)$ and obeying a certain set of
  second-order nonlinear differential equations,
\beq\label{0.1}
     \frac{d}{dz} \left( \frac{ z}{H_s} \frac{d}{dz} H_s \right) =
        \frac{1}{4}  B_s \prod_{s' \in S}  H_{s'}^{- A_{s s'}},
\eeq
  with the boundary conditions
\beq\label{0.2}
       H_{s}(+ 0) = 1,
\eeq
  $s \in S$ ($S$ is a non-empty set). In (\ref{0.1}), all $B_s \neq 0$ are
  constants and  $(A_{s s'})$ is the so-called ``quasi-Cartan'' matrix
  ($A_{ss} = 2$) coinciding with the Cartan matrix when intersections are
  related to Lie algebras.

  In \cite{Iflux}, the following {\bf hypothesis} was suggested: the
  solutions to \eqs (\ref{0.1}), (\ref{0.2}) (if they exist) are polynomials
  when the intersection rules correspond to semisimple Lie algebras.  As
  pointed in \cite{Iflux}, this hypothesis could be readily verified for
  the Lie algebras $A_m$, $C_{m+1}$, $m = 1,2, \ldots$, as was done for
  black-brane solutions from \cite{IMp1,IMp3}.

  In \cite{Iflux}, explicit formulae for solutions corresponding to the Lie
  algebras $A_1 \oplus \ldots \oplus A_1$ and $A_2$ were presented.

  In this paper, we consider generalized ``flux-S-brane'' solutions
  depending on the parameter $w = \pm 1$. For $w = +1$, these solutions are
  coinciding with flux-brane solutions from \cite{Iflux}. For $w = -1$, they
  describe special S-brane solutions with Ricci-flat factor spaces. (For
  general S-brane configurations see \cite{Isbr} and references therein.)
  Here we present new solutions with polynomials $H_s(z)$ related to
  the algebras $C_{2}$ and $G_2$.

  The $S$-brane solutions corresponding to the Lie algebras $A_2$, $C_2$ and
  $G_2$, for special choices of the parameters, may describe an
  accelerating expansion of ``our'' 3-dimensional space with small enough
  variations of the effective gravitational constant
  \cite{IMacg} (for the case of the $A_2$ algebra see also \cite{AIKMa}).

\section{Flux- and $S$-brane solutions with general intersection rules }

\subsection{The model}

  We consider a  model governed by the action
\bearr\label{1.1}
    S=\int d^Dx \sqrt{|g|}\biggl\{R[g]
        -h_{\alpha\beta}g^{MN}\d_M\varphi^\alpha  \d_N\varphi^\beta
\nnn \cm\cm
    - \sum_{a\in\tri}\frac{\theta_a}{N_a!}
        \exp[2\lambda_a(\varphi)](F^a)^2\biggr\},
\ear
  where $g=g_{MN}(x)dx^M\otimes dx^N$ is the metric, $\varphi =
  (\varphi^\alpha) \in \R^l$ is a vector of scalar fields,
  $(h_{\alpha\beta})$ is a  constant symmetric non-degenerate $l\times l$
  matrix ($l\in \N$), $\theta_a = \pm 1$,
\beq\label{1.2a}
    F^a =  dA^a  =  \frac{1}{N_a!} F^a_{M_1 \ldots M_{n_a}}
        dz^{M_1} \wedge \ldots \wedge dz^{M_{n_a}}
\eeq
  is an $N_a$-form ($n_a\ge 1$),  $\lambda_a$ is a 1-form on $\R^l$:
  $\lambda_a(\varphi)=\lambda_{a \alpha}\varphi^\alpha$, $a \in \tri$,
  $\alpha=1,\dots,l$.  In (\ref{1.1}), we denote $|g| = |\det (g_{MN})|$,
\beq\label{1.3a}
    (F^a)^2_g  =
    F^a_{M_1 \ldots M_{n_a}} F^a_{N_1 \ldots N_{n_a}}
        g^{M_1 N_1} \ldots g^{M_{n_a} N_{n_a}},
\eeq
  $a \in \tri$, where $\tri$ is some finite set.

\subsection{``Flux-S-brane'' solutions}

  Let us consider a family of exact solutions to the field equations
  corresponding to the action (\ref{1.1}) and depending on one variable
  $\rho$. These solutions are defined on the manifold
\beq\label{1.2}
    M = (0, + \infty)  \times M_1 \times M_2 \times \ldots \times M_{n},
\eeq
  where $M_1$ is a one-dimensional manifold. The solutions read
\bearr   \label{2.30}
        g= \Bigl(\prod_{s \in S} H_s^{2 h_s d(I_s)/(D-2)} \Bigr)
            \biggl\{ w d\rho \otimes d \rho
\nnn \
  +  \Bigl(\prod_{s \in S} H_s^{-2 h_s} \Bigr) \rho^2 g^1 +
        \sum_{i = 2}^{n} \Bigl(\prod_{s\in S}
            H_s^{-2 h_s \delta_{iI_s}} \Bigr) g^i  \biggr\},
\yyy  \label{2.31}
  \exp(\varphi^\alpha)=\prod_{s\in S} H_s^{h_s \chi_s \lambda_{a_s}^\alpha},
\yyy  \label{2.32a}
   F^a= \sum_{s \in S} \delta^a_{a_s} {\cal F}^{s},
\ear
  where
\bearr\label{2.32}
    {\cal F}^s= - Q_s \left( \prod_{s' \in S}  H_{s'}^{- A_{ss'}} \right)
        \rho d\rho  \wedge \tau(I_s),
\nnn \inch
        s\in S_e,
\\ \lal \label{2.33}
    {\cal F}^s= Q_s \tau(\bar I_s), \cm s\in S_m.
\ear

  The functions $H_s(z) > 0$, $z = \rho^2$ obey \eqs (\ref{0.1})
  with the boundary conditions (\ref{0.2}).

  In  (\ref{2.30}),  $g^i=g_{m_i n_i}^i(y_i) dy_i^{m_i}\otimes dy_i^{n_i}$
  is a Ricci-flat metric on $M_{i}$, $i=  1,\ldots,n$,
\beq\label{1.11}
      \delta_{iI}=  \sum_{j\in I} \delta_{ij}
\eeq
  is the indicator of $i$ belonging to $I$: $\delta_{iI}=  1$ for
  $i\in I$ and $\delta_{iI}=  0$ otherwise.

  The brane set $S$ is, by definition, a union of two sets:
\beq\label{1.6}
    S=  S_e \cup S_m, \quad
    S_v=  \cup_{a\in\tri}\{a\}\times\{v\}\times\Omega_{a,v},
\eeq
  $v=  e,m$ and $\Omega_{a,e}, \Omega_{a,m} \subset \Omega$,
  where $\Omega =   \Omega(n)$  is the set of all non-empty subsets of
  $\{ 1, \ldots,n \}$. Any brane index $s \in S$ has the form
\beq\label{1.7}
     s =   (a_s,v_s, I_s),
\eeq
  where $a_s \in \tri$ is the colour index, $v_s =  e,m$ is the
  electro-magnetic index, and the set $I_s \in \Omega_{a_s,v_s}$ describes
  the location of the brane worldvolume.

  The sets $S_e$ and $S_m$ define electric and magnetic branes,
  respectively. In (\ref{2.31}),
\beq\label{1.8}
     \chi_s  =   +1, -1
\eeq
  for $s \in S_e,\ S_m$, respectively. In (\ref{2.32a}), the forms
  (\ref{2.32}) correspond to electric branes and the forms (\ref{2.33})
  to magnetic branes; $Q_s \neq 0$, $s \in S$. In (\ref{2.33}) and in what
  follows,
\beq \label{1.13a}
  \bar I \equiv I_0 \setminus I, \qquad  I_0 = \{1,\ldots,n \}.
\eeq

  All manifolds $M_{i}$ are assumed to be oriented and connected, and
  the volume $d_i$-forms
\beq\label{1.12}
  \tau_i \equiv \sqrt{|g^i(y_i)|} \ dy_i^{1} \wedge \ldots \wedge dy_i^{d_i},
\eeq
  and the parameters
\beq\label{1.12a}
     \eps(i)  \equiv \sign \det (g^i_{m_i n_i}) = \pm 1
\eeq
  are well defined for all $i= 1,\ldots,n$. Here $d_{i} = \dim M_{i}$, $i =
  1, \ldots, n$, $D = 1 + \sum_{i = 1}^{n} d_{i}$. For any $I =\{ i_1,
  \ldots, i_k \} \in \Omega$, $i_1 < \ldots < i_k$, we denote
\bear\label{1.13}
    \tau(I) \eqv \tau_{i_1}  \wedge \ldots  \wedge \tau_{i_k},
\yy \label{1.14}
    M(I) \eqv M_{i_1}  \times  \ldots \times M_{i_k},
\yy  \label{1.15}
       d(I) \eqv {\rm dim } M(I) =  \sum_{i \in I} d_i,
\\
    \eps(I) \eqv \eps(i_1) \ldots \eps(i_k).
\ear
  $M(I_s)$ is isomorphic to a brane worldvolume (see (\ref{1.7})).

  The parameters  $h_s$ appearing in the solution satisfy the relations
\beq\label{1.16}
    h_s = K_s^{-1}, \qquad  K_s = B_{s s},
\eeq
  where
\bearr\label{1.17}  \nhq
     B_{ss'} \equiv    d(I_s\cap I_{s'})+\frac{d(I_s)d(I_{s'})}{2-D}+
    \chi_s \chi_{s'} \lambda_{a_s \alpha} \lambda_{a_{s'} \beta}
             h^{\alpha\beta},
\nnn
\ear
  $s, s' \in S$, with $(h^{\alpha\beta}) = (h_{\alpha\beta})^{-1}$. In
  (\ref{2.31}), $\lambda_{a_s}^{\alpha} = h^{\alpha\beta}\lambda_{a_s\beta}$.

  We will assume that
\beq\label{1.17a}
    ({\bf i}) \qquad B_{ss} \neq 0,
\eeq
  for all $s \in S$, and
\beq\label{1.18b}
       ({\bf ii}) \qquad \det (B_{s s'}) \neq 0,
\eeq i.e.
  the matrix $(B_{ss'})$ is nondegenerate. In (\ref{2.32}), there appears
  another nondegenerate matrix (the so-called ``quasi-Cartan'' matrix)
\beq\label{1.18}
    (A_{ss'}) = \left( 2 B_{s s'}/B_{s' s'} \right).
\eeq
  In (\ref{0.1}),
\beq\label{1.21}
      B_s = \eps_s K_s Q_s^2, \cm s\in S,
\eeq
  where
\beq\label{1.22}
      \eps_s=(-\eps[g])^{(1-\chi_s)/2}\eps(I_s) \theta_{a_s},
\eeq
  $s\in S$, $\eps[g] \equiv \sign\det (g_{MN})$.

  More explicitly, (\ref{1.22}) reads: $\eps_s= \eps(I_s) \theta_{a_s}$ for
  $v_s = e$ and $\eps_s = -\eps[g] \eps(I_s) \theta_{a_s}$, for $v_s = m$.

  Due to (\ref{2.32}) and  (\ref{2.33}), the brane worldvolume dimension
  $d(I_s)$ is determined as
\beq\label{1.16a}
      d(I_s)=  N_{a_s}-1,   \cm   d(I_s) = D- N_{a_s} -1,
\eeq
  for $s \in S_e,\ S_m$, respectively. For an $Sp$-brane: $p =p_s =d(I_s)-1$.

\medskip\noi
  {\bf Restrictions on brane configurations.} The solutions presented above
  are valid if two restrictions on the sets of branes are satisfied. These
  restrictions guarantee a block-diagonal form of the energy-momentum
  tensor and the existence of the sigma-model representation (without
  additional constraints) \cite{IMs,IMtop}. These restrictions are:
\beq\label{2.2.2a}
     {\bf (R1)} \qquad d(I \cap J) \leq d(I)  - 2
\eeq
  for any $I,J \in\Omega_{a,v}$, $a\in\tri$, $v= e,m$ (and here $d(I)
  = d(J)$).

\beq\label{2.2.3a}
    {\bf (R2)} \qquad d(I \cap J) \neq  0,
\eeq
  for $I \in \Omega_{a,e}$ and $J \in \Omega_{a,m}$, $a \in \tri$.

  In the cylindrically symmetric case,
\beq\label{2.40}
     M_1 = S^1, \qquad g^1 = d \phi \otimes d \phi,
\eeq
  $0 < \phi < 2 \pi$ and $w = +1$, we get the family of composite fluxbrane
  solutions from \cite{Iflux}.

\section{Conjecture on a polynomial structure of $H_s$ for  Lie algebras}

  In what follows, we consider the case
\bearr\label{eps}
        \eps_s > 0,
\\ \lal   \label{K}
        K_s > 0.
\ear
  In this case all $B_s > 0$. \eq (\ref{eps}) is satisfied when all
  $\theta_a > 0$, $\eps[g] = -1$ (e.g., when the metric $g$ has a
  pseudo-Euclidean signature $(-,+,...,+)$)  and
\beq\label{eps1}
      \eps(I_s) = +1
\eeq
  for all $s \in S$. (The relation (\ref{eps1})  takes place for $S$-brane
  and fluxbrane solutions.) The second relation (\ref{K}) takes place when
  all $d(I_s) < D-2$ and the matrix $(h_{\alpha\beta})$ is positive-definite
  (i.e., there are no phantom scalar fields).

  Let us consider the second-order differential equations  (\ref{0.1})
  with the boundary conditions (\ref{0.2}) for the functions $H_s(z) > 0$,
  $s \in S$. We will be interested in analytical solutions to \eqs
  (\ref{0.1}) in some disc $|z| < L$:
\beq\label{3.3}
      H_{s}(z) = 1 + \sum_{k = 1}^{\infty} P_s^{(k)} z^k,
\eeq
  where $P_s^{(k)}$ are constants, $s \in S$. Substitution of (\ref{3.3})
  into (\ref{0.1}) gives an infinite chain of relations for the parameters
  $P_s^{(k)}$  and $B_s$.  The first relation in this chain
\beq\label{3.5a}
    P_s  \equiv  P_s^{(1)} = \frac{1}{4} B_s = \frac{1}{4}K_sQ^2_s,
\eeq
  $s \in S$, corresponds to the $z^0$-term in the decomposition of
  (\ref{0.1}).

  It can by shown that, for analytic functions $H_s(z)$, $s \in S$
  (\ref{3.3}) ($z = \rho^2$), the metric (\ref{2.30}) is regular at
  $\rho = 0$ for $w = + 1$, i.e. in the fluxbrane case.

  Let $(A_{s s'})$ be a Cartan matrix of a finite-dimensional semisimple Lie
  algebra $\cal G$.

  It has been conjectured in \cite{Iflux} that there exist polynomial
  solutions to \eqs (\ref{0.1}), (\ref{0.2}), having the form
\beq\label{3.12}
     H_{s}(z) = 1 + \sum_{k = 1}^{n_s} P_s^{(k)} z^k,
\eeq
  where $P_s^{(k)}$ are constants, $k = 1,\ldots, n_s$. Here, $P_s^{(n_s)}
  \neq 0$,  $s \in S$, and
\beq\label{3.11}
      n_s = 2 \sum_{s' \in S} A^{s s'}.
\eeq
  The integers $n_s$ are components of the so-called twice dual Weyl vector
  in the basis of simple roots \cite{FS}.

\subsection{Examples of solutions for rank-2 Lie algebras}

  Consider configurations with two branes, i.e., $S = \{s_1, s_2 \}$.

\subsubsection{Solutions in the $A_1 \oplus   A_1$ case}

  The simplest example occurs in so-called ``orthogonal'' case, when
  $(A_{s s'}) = \diag (2,2)$ is the Cartan matrix of the semisimple Lie
  algebra $A_1 \oplus A_1$, where $A_1 = sl(2)$. We get \cite{Iflux}
\beq\label{3.5}
      H_{s}(z) = 1 + P_s z,
\eeq
  with $P_s \neq 0$ satisfying (\ref{3.5a}).

\subsubsection{Solutions in the $A_2$ case}

  For the Lie algebra $A_2 = sl(3)$ with the Cartan matrix
\beq\label{4.5a}
    (A_{ss'})= \pmatrix{      2 & -1\cr
                     -1 &  2\cr },
\eeq
  we have \cite{Iflux} $n_1 = n_2 =2$:
\beq\label{4.6}
    H_{s} = 1 + P_s z + P_s^{(2)} z^{2},
\eeq
  where, here and in what follows, $P_s$ obey \eq (\ref{3.5a}), and
\beq\label{4.7}
       P_s^{(2)}  =  \frac 14 P_1 P_2.
\eeq

\subsubsection{Solutions for the Lie algebra $C_2$}

  For the Lie algebra $C_2 = so(5)$ with the Cartan matrix
\beq\label{4.5c}
    (A_{ss'})=  \pmatrix{       2 & -1 \cr
                       -2 &  2 \cr   },
\eeq
  we get from  (\ref{3.11}): $n_1 = 3$ and $n_2 = 4$. For the moduli
  functions we obtain
\bear\label{4.6c1} \nhq
    H_1 \eql 1+P_1 z+ \frac{1}{4} P_1 P_2 z^2 + \frac{1}{36}P_1^2P_2 z^3,
\\ \label{4.6c2}   \nhq
        H_2 \eql 1+ P_2 z+ \frac{1}{2}P_1 P_2 z^2
      +\frac{1}{9}P_1^2 P_2z^3 +\frac{1}{144}P_1^2 P_2^2 z^4.
\nnn
\ear

\subsubsection{Solutions for the Lie algebra $G_2$}

  Consider now the exceptional Lie algebra $G_2$ with the Cartan matrix
\beq\label{4.5g}
    (A_{ss'}) = \pmatrix{      2 & -1 \cr
                      -3 &  2 \cr }.
\eeq
  We get from  (\ref{3.11})  $n_1 = 6$ and $n_2 = 10$. Calculations
  (using MATHEMATICA and MAPLE) give:
\bear   \label{4.6d1}  \nq
    H_1 \eql 1+P_1 z+ \frac{1}{4}P_1P_2 z^2 +\frac{1}{18} P_1^2
             P_2 z^3 + \frac{1}{144}P_1^3 P_2 z^4
\nnn \ \
  + \frac{1}{3600}P_1^3 P_2^2 z^5 + \frac{1}{129600}P_1^4 P_2^2 z^6 ,
\\ \label{4.6d2}  \nq
        H_2 \eql 1+ P_2 z+ \frac{3}{4} P_1 P_2 z^2 +\frac{1}{3}P_1^2 P_2 z^3
\nnn \ \
  +\frac{1}{16} P_1^2 P_2(\frac{1}{3}P_2+P_1) z^4
                +\frac{7}{600} P_1^3 P_2^2 z^5
\nnn \ \
   + \frac{1}{64} P_1^3 P_2^2\biggl(\frac{1}{25}P_2
   +\frac{1}{27}P_1\biggr)z^6   +\frac{1}{10800}P_1^4 P_2^3  z^7
\nnn \ \
   +\frac{1}{172800}P_1^5 P_2^3 z^8 +  \frac{1}{4665600} P_1^6 P_2^3 z^9
\nnn \ \
      + \frac{1}{466560000}P_1^6 P_2^4 z^{10}.
\ear

  The intersection rules are given by \eqs (\ref{1.18}).

  We would like to outline some useful relations for the big numbers
  appearing in the denominators of the polynomial coefficients:
  $1296 = 6^4$, $1728 =  3 (24^2)$, $46656 = 6^6$.

  There are at least two ways of calculating the coefficients
  $P_s^{(k)}$ of the polynomials. The first one (performed by MAPLE)
  consists in a straightforward substitution of the polynomials (\ref{3.12})
  into the ``master'' equations (\ref{0.1}). The second
  one (carried out with MATHEMATICA) uses recurrent relations for the
  coefficients $P_s^{(k +1)}$ as functions of other coefficients $P_s^{(1)},
  \ldots,\ P_s^{(k)}$. These recurrent relations were obtained analytically by
  simply decomposing  the ``master'' equations (\ref{0.1}) into a power
  series in the parameters $z$ \cite{Ir}.

\section{Conclusions}

  We have presented explicit formulae for fluxbrane and S-brane solutions
  governed by polynomials which correspond to Lie algebras: $A_1 \oplus
  A_1$, $A_2$, $C_2$ and $G_2$. The pairs of moduli functions $(H_1,\ H_2)$
  in these solutions are polynomials of degrees: $(1,\ 1)$, $(2,\ 2)$,
  $(3,\ 4)$ and $(6,\ 10)$, in agreement with a conjecture from
  Ref.\,\cite{Iflux}. The general S-brane solutions presented here, governed
  by polynomials, are new.  The fluxbrane solutions related to Lie
  algebras $C_2$ and $G_2$ are new as well.

\Acknow
{This work was supported in part by the Russian Foundation for
Basic Research grant Nr. 05-02-17478 and by DFG grant  Nr. 436 RUS
113/807/0-1. }

\end{document}